%% file: paper.tex
\documentclass[11pt,a4paper]{article}
\pdfoutput=1
\usepackage[hyperref]{conll2018}
\usepackage{times}
\usepackage{latexsym}
\newcommand*\cleartoleftpage{%
  \clearpage
  \ifodd\value{page}\hbox{}\newpage\fi
}

\usepackage{times}
\usepackage{rotating}
\usepackage{latexsym}
\usepackage{amsmath}

\usepackage{mathptmx}
\usepackage{mathtools,xparse}
\usepackage{amssymb}
\usepackage{caption}
\usepackage{graphicx}
\usepackage{epstopdf}
\usepackage{color}
\usepackage{url}
\usepackage{tabularx}
\usepackage{xfrac}
\usepackage{array}
\usepackage[labelfont=bf]{caption}
\usepackage{color}
\usepackage{subcaption}
\usepackage{float}
\usepackage{multirow}
\usepackage{enumitem}
\usepackage{algorithm,algorithmic}
\usepackage{hhline}
 \usepackage{tikz}
\usepackage{booktabs}
\usepackage{adjustbox}
\let\subparagraph\relax
\usepackage{titlesec}
\usepackage{hyperref}
\definecolor{darkblue}{rgb}{0,0,.7}
\definecolor{armygreen}{rgb}{0.29, 0.33, 0.13}
\definecolor{cobalt}{rgb}{0.0, 0.28, 0.67}
\hypersetup{
     colorlinks   = true,
     citecolor    = armygreen,
     urlcolor     = darkblue,
     linkcolor    = darkblue
     }
\aclfinalcopy
\title{A Trio Neural Model for Dynamic Entity Relatedness Ranking}

\author{Tu Nguyen \\
  L3S Research Center \\
  {\tt tunguyen@L3S.de} \\\And
  Tuan Tran \\
  Robert Bosch GmbH \\
  {\tt \small{anhtuan.tran2@de.bosch.com}}\\\And
  Wolfgang Nejdl \\
  L3S Research Center \\
  {\tt nejdl@L3S.de} \\}
\makeatletter
\renewcommand{\@date}{}
\makeatother
\begin{document}
%
%

%
%


\maketitle
\begin{abstract}
Measuring entity relatedness is a fundamental task for many natural language processing and information retrieval applications. Prior work often studies entity relatedness in static settings and an unsupervised manner. However, entities in real-world are often involved in many different relationships, consequently entity-relations are very dynamic over time. In this work, we propose a neural network-based approach for \emph{dynamic} entity relatedness, leveraging the \textit{collective attention} as supervision. Our model is capable of learning rich and different entity representations in a joint framework. Through extensive
experiments on large-scale datasets,
we demonstrate that our method achieves better results than competitive
baselines.
\end{abstract}

\input{intro}
\input{relatedwork}
\input{preliminaries}
\input{methodology}

\input{experiment}
\input{conclusion}
\input{acknowledgements}

\bibliographystyle{acl_natbib}
\bibliography{pastbib}

\end{document}

%% file: intro.tex
\section{Introduction}
\label{sec:intro}

Measuring semantic relatedness between entities is an inherent component in many text mining applications. In search and recommendation, the ability to suggest most related entities to the entity-bearing query has become a standard feature of popular Web search engines~\cite{blanco2013entity}. In natural language processing, entity relatedness is an important factor for various tasks, such as entity linking~\cite{hoffart2012kore} or word sense disambiguation~\cite{moro2014entity}.

 	
However, prior work on semantic relatedness often neglects the time dimension and consider entities and their relationships as static. In practice, many entities are highly ephemeral~\cite{jiang2016towards}, and users seeking information related to those entities would like to see fresh information. For example, users looking up the entity \textsf{Taylor Lautner} during 2008--2012 might want to be recommended with entities such as \textsf{The Twilight Saga}, due to Lautner's well-known performance in the film series; however the same query in August 2016 should be served with entities related to his appearances in more recent films such as ``Scream Queens'', ``Run the Tide''. 
In addition, much of previous work resorts to deriving semantic relatedness from co-occurence-based computations or heuristic functions without direct optimization to the final goal. 
We believe that desirable framework should see entity semantic relatedness as not separate but an integral part of the process, for instance in a supervised manner. 

In this work, we address the problem of \textbf{\emph{entity relatedness ranking}}, that is, designing the semantic relatedness models that are optimized for ranking systems such as top-$k$ entity retrieval or recommendation. In this setting, the goal is not to quantify the semantic relatedness between two entities based on their occurrences in the data, but to optimize the partial order of the related entities in the top positions. This problem differs from traditional entity ranking~\cite{kang2015learning} in that the entity rankings are driven by user queries and are optimized to their (ad-hoc) information needs, while entity relatedness ranking also aims to uncover the meanings of the the relatedness from the data. In other words, while conventional entity semantic relatedness learns from data (editors or content providers' perspectives), and entity ranking learns from the user's perspective, the entity relatedness ranking takes the trade-off between these views. Such a hybrid approach can benefit applications such as exploratory entity search~\cite{miliaraki2015selena}, where users have a specific goal in mind, but at the same time are opened to other related entities.  

\begin{figure}[t]
\centering
		\includegraphics[width=0.9\columnwidth]{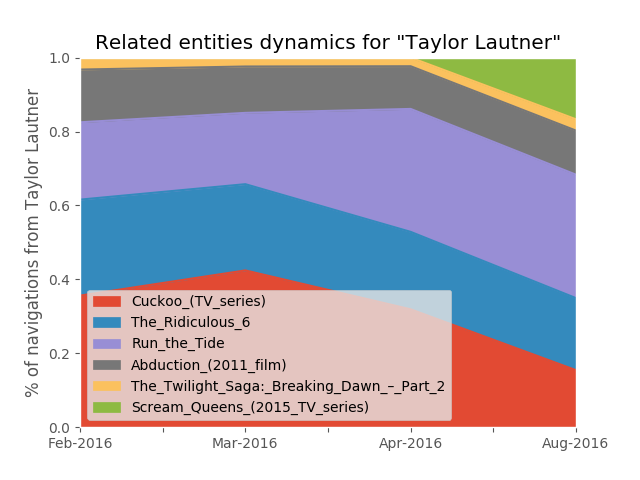}\vspace{-0.2cm}
		\caption{The dynamics of collective attention for related entities of \textsf{Taylor Lautner} in 2016.}		
		\label{fig:showcase1}
\end{figure}

We also tackle the issue of \emph{dynamic ranking} and design the supervised-learning model that takes into account the temporal contexts of entities, and proposes to leverage \emph{collective attention} from public sources. As an illustration, when one looks into the Wikipedia page of \textsf{Taylor Lautner}, each navigation to other Wikipedia pages indicates the user interest in the corresponding target entity given her initial interest in Lautner. Collectively, the navigation traffic observed over time is a good proxy to the shift of public attention to the entity (Figure~\ref{fig:showcase1}). 

In addition, while previous work mainly focuses on one aspect of the entities such as textual profiles
 or linking graphs
 , we propose a \emph{trio neural} model that learns the low level representations of entities from three different aspects: Content, structures and time aspects. For the time aspect, we propose a convolutional model to \textit{ embed} and \textit{attend} to local patterns of the past temporal signals in the Euclidean space. 
 Experiments show that our trio model outperforms traditional approaches in ranking correlation and recommendation tasks. 
Our contributions are summarized as follows.

\begin{itemize}
\setlength\itemsep{0em}
\item We formulate \textit{dynamic entity relatedness ranking} and
optimize directly for time-sensitive \emph{pairwise ordering} rather
than static similarity scores.

\item We introduce a temporal convolutional module with a
monotonic time-decay weighting scheme, enabling the model to embed
temporal signals and emphasize recency-dependent local patterns.

\item We propose a trio neural ranking framework that jointly models
content-, graph-, and time-based views of entities,  yielding jmutually informative representations tailored for ranking.
\end{itemize}
\vspace{-0.2cm}

%% file: relatedwork.tex
\section{Related Work}
\label{sec:related}
\subsection{Entity Relatedness and Recommendation}
Most of existing semantic relatedness measures (e.g. derived from Wikipedia) can be divided into the following two major types: (1) \textit{text}-based, (2) \textit{graph}-based. For the first, traditional methods mainly focus on a high-dimensional semantic space based on occurrences of words (~\newcite{Gabrilovich:2007:CSR:1625275.1625535,gabrilovich2009wikipedia}) or concepts (~\newcite{aggarwal2014wikipedia}). 
In recent years, embedding methods that learn low-dimensional word representations have been proposed. ~\newcite{hu2015entity} leverages entity embedding on knowledge graphs to better learn the distributional semantics.~\newcite{Ni:2016:SDR:2835776.2835801} use an adapted version of Word2Vec, where each entity in a Wikipedia page is considered as a term. 
For the \textit{graph}-based approaches, these measures usually take advantage of the hyperlink structure of entity graph~\cite{witten2008effective,guo2014robust}. Recent graph embedding techniques (e.g., DeepWalk~\cite{perozzi2014deepwalk}) have not been directly used for entity relatedness in Wikipedia, yet its performance is studied and shown very competitive in recent related work~\cite{zhao2015representation,Ponza:2017:TFC:3132847.3132890}. 

Entity relatedness is also studied in connection with the entity recommendation task. The Spark~\cite{blanco2013entity} system firstly introduced the task for Web search, ~\newcite{yu2014building,zhang2016collaborative} exploit user click logs and entity pane logs for global and personalized entity recommendation. However, these approaches are optimized to user information needs, and also does not target the \textit{global} and \textit{temporal} dimension. Recently, ~\newcite{zhang2016probabilistic,tran2017beyond} proposed time-aware \textit{probabilistic} approaches that combine `static' entity relatedness with temporal factors from different sources.~\newcite{DBLP:conf/esws/NguyenKN18} studied the task of time-aware ranking for entity aspects and propose an ensemble model to address the sub-features competing problem. 


\subsection{Neural Network Models}
\textbf{Neural Ranking.} Deep neural ranking among IR and NLP can be generally divided into two groups:  representation-focused and interaction-focused models. The \textit{representation-focused} approach~\cite{huang2013learning} independently learns a representation for each ranking element (e.g., query and document) and then employ a similarity function. On the other hand, the \textit{interaction-focused} models are designed based on the early interactions between the ranking pairs as the input of network. For instance, ~\newcite{lu2013deep,guo2016deep} build interactions (i.e., local matching signals) between two pieces of text and trains a feed-forward network for computing the matching score. 
This enables the model to capture various interactions between ranking elements, while with former, the model has only the chance of isolated observation of input elements.


\textbf{Attention networks.} In recent years, attention-based NN architectures, which learn to focus their ``attention'' to specific parts of the input, have shown promising results on various NLP tasks. For most cases, attentions are applied on \textit{sequential models} to capture \textit{global} context ~\cite{luong2015effective}. An attention mechanism often relies on a \textit{context} vector that facilitates outputting a ``summary'' over all (deterministic soft) or a sample (stochastic hard) of input states.  
Recent work proposed a CNN with attention-based framework to model \textit{local} context representations of textual pairs ~\cite{yin2016abcnn}, or to combine with LSTM to model \textit{time-series} data~\cite{ordonez2016deep,lin2017hybrid} for classification and trend prediction tasks.
\vspace{-0.1cm}

%% file: preliminaries.tex
\section{Problem}
\vspace{-0.1cm}
\subsection{Preliminaries}
\label{sec:prel}
\vspace{-0.1cm}
We denote as named entities any real-world objects 
registered in a database. Each entity has a textual document (e.g. content of a home page), and a sequence of references to other entities (e.g., obtained from semantic annotations), called the entity ~\textit{link profile}. All link profiles constitute an entity linking graph. 
In addition, two types of information are included to form the entity collective attention. 

\textbf{Temporal signals.}  Each entity can be associated with a number of properties such as view counts, content edits, etc. Given an entity $e$ and a time point $n$, given $D$ properties, the temporal signals set, in the form of a (univariate or multivariate) \textit{time series} $X\in \mathbf{R}^{D \times T}$ consists of $T$ real-valued vector $x_{n-T}, \cdots, x_{n-1}$ , where $x_{t} \in \mathbf{R}^{D}$ captures the past signals of $e$ at time point $t$. 


\textbf{Entity Navigation.}
In many systems, the user navigation between two entities is captured, e.g., search engines can log the total click-through of documents of the target entity presented in search results of a query involving the source entity. Following learning to rank approaches~\cite{kang2015learning}, we use this information as the ground truth in our supervised models. 
Given two entities $e_1, e_2$, the navigation signal from $e_1$ to $e_2$ at time point $t$ is denoted by $y_{\{e_{1},e_{2}\}}^{t}$. 



\subsection{Problem Definition}
\label{subsec:problem}

We consider the task of quantifying and ranking the semantic relatedness
between entities as it evolves over time. Rather than assuming a single
static relatedness function, we allow the notion of relatedness to vary
with time and model it as a family of functions
$F = \{f_t\}_{t \in \mathcal{T}}$, where each $f_t$ reflects the
relations observed at time $t$.

\paragraph{Dynamic Entity Relatedness.}
For any time point $t$ and any pair of entities $(e_s, e_t)$ consisting
of a \emph{source} entity $e_s$ and a \emph{target} entity $e_t$, the
dynamic relatedness score is given by a function
\[
f_t : \mathcal{E} \times \mathcal{E} \rightarrow \mathbb{R}_{\ge 0}.
\]
We require each $f_t$ to satisfy the following properties:
\begin{itemize}
\item \textbf{Asymmetry:} \; $f_t(e_i, e_j)$ need not equal
      $f_t(e_j, e_i)$, reflecting directional relationships.
\item \textbf{Non-negativity:} \; $f_t(e_i, e_j) \ge 0$ for all entity
      pairs.
\item \textbf{Indiscernibility:} \; $e_i = e_j$ implies
      $f_t(e_i, e_j) = 1$, corresponding to maximal self-relatedness.
\end{itemize}
We make no assumptions of symmetry, transitivity, or metric structure,
allowing $f_t$ to capture complex and time-varying semantic interactions.

\paragraph{Dynamic Entity Relatedness Ranking.}
Given a source entity $e_s$ and a time point $t$, the goal is to rank a
set of candidate target entities $\{e_1, \ldots, e_n\}$ according to
their dynamic relatedness scores
$f_t(e_s, e_1), \ldots, f_t(e_s, e_n)$.  
The output is an ordering that reflects the relative strength of
entity–entity associations at time $t$.

%% file: methodology.tex
\section{Approach Overview}
\label{sec:approach}

\subsection{Datasets and Their Dynamics}
In this work we use Wikipedia data as the case study for our entity relatedness ranking problem due to its rich knowledge and dynamic nature. It is worth noting that despite experimenting on Wikipedia, our framework is universal can be applied to other sources of entity with available temporal signals and entity navigation. We use Wikipedia pages to represent entities and page views as the temporal signals (details in section~\ref{ref:exp:dataset}).
\input{clickstream}

\begin{figure}[t]
\centering
		\includegraphics[width=0.6\columnwidth]{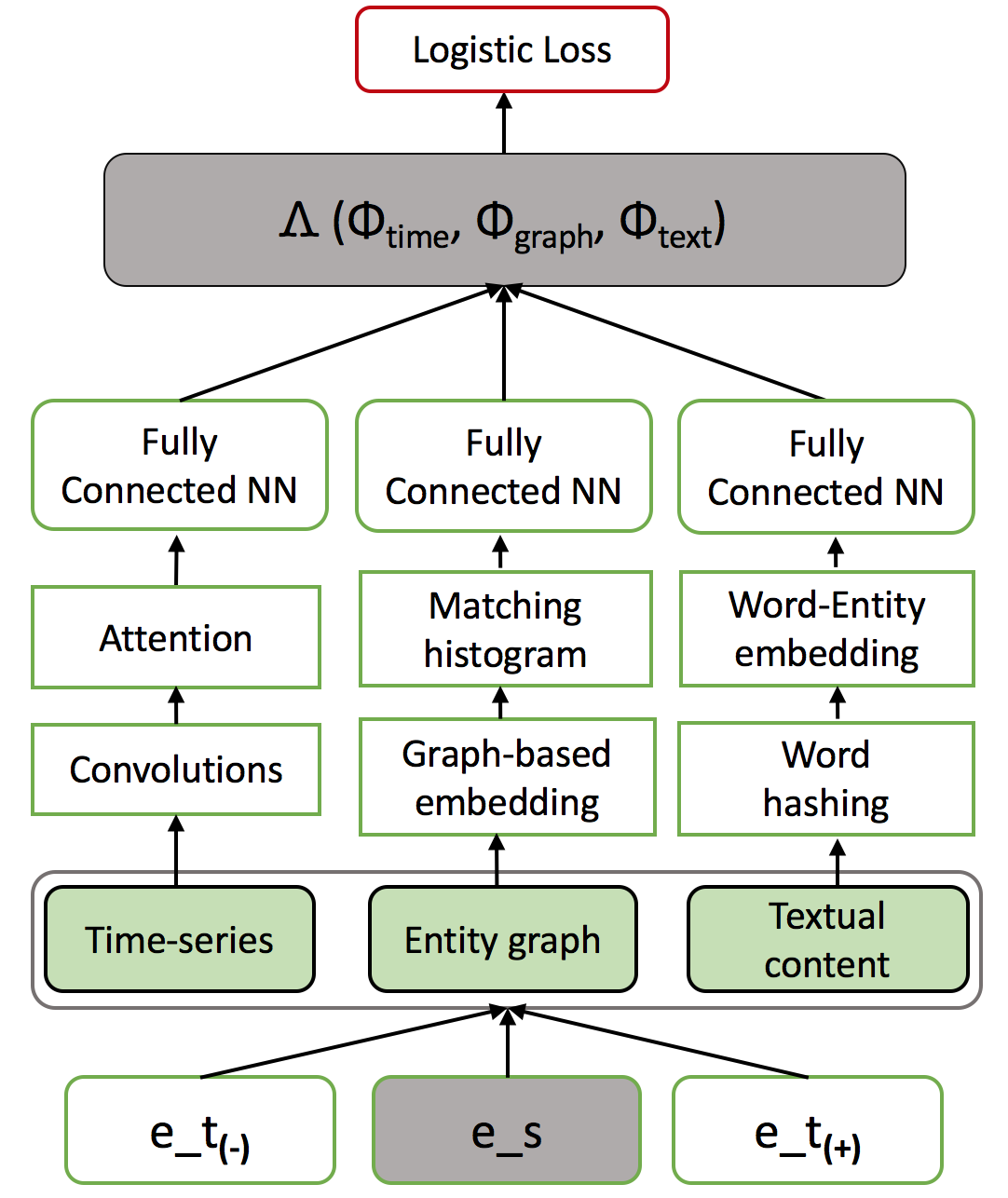}\vspace{-0.1cm}
		\caption{The trio neural model for entity ranking.}		
		\label{fig:showcase2}
\end{figure}

Figure~\ref{fig:showcase2} shows the overall architecture of our framework, which consists of three major components: \textit{time}-, \textit{graph}- and \textit{content}-based networks. Each component can be considered as a separate sub-ranking network. Each network accepts a tuple of three elements/representations as an input in a \textit{pair-wise} fashion, i.e., the source entity $e_{s}$, the target entity $e_{t}$ with higher rank (denoted as $e_{(+)}$) and the one with lower rank (denoted as $e_{(-)}$). For the \textit{content} network, each element is a
sequence of terms, coming from entity textual representation. 
For the \textit{graph} network, we learn the embeddings from the entity linking graph. For the \textit{time} network, we propose a new convolutional model learning from the entity temporal signals. More detailed are described as follows. 

\vspace{-0.1cm}

\subsection{Neural Ranking Model Overview}

Entity relatedness can in principle be modeled using a point-wise
approach that directly predicts a scalar score for each entity pair.
However, navigation data is highly skewed, and supervision from
long-tail interactions is often noisy.  
Instead of learning a fully calibrated scoring function, we therefore
adopt a pair-wise ranking strategy, which focuses on comparing candidate
entities relative to one another.  
Pair-wise methods have the advantage of preserving partial orders of the
underlying relatedness functions $f_t$ even when $f_t$ is not globally
transitive \cite{NIPS2012_4811}, making them well-suited for dynamic
entity ranking.

Our architecture follows an interaction-based design in which the model
directly processes the triplet $(e_s, e_{(+)}, e_{(-)})$ to learn their
relative compatibility.  
In contrast to Siamese or representation-based models
\cite{chopra2005learning}, our networks do \emph{not} share parameters
across branches.  
Parameter sharing would implicitly enforce symmetry
$f_t(e_s, e) = f_t(e, e_s)$, contradicting the inherently asymmetric
nature of the relatedness function (Section~\ref{subsec:problem}).

Each branch of the model is a feed-forward network $\psi$ with input
$z_0$, hidden layers $z_1,\ldots,z_{n-1}$, and output $z_n$, where
\[
z_i = \sigma(\mathbf{W}_i z_{i-1} + \mathbf{b}_i),
\]
and $\sigma$ is a nonlinear activation such as ReLU.  
Under the trio setup, the overall pair-wise score is the sum of the
outputs from the temporal, graph, and content networks:
\[
\phi(e_s, e_{(+)}, e_{(-)}) 
=
\phi_{\mathrm{time}}
+
\phi_{\mathrm{graph}}
+
\phi_{\mathrm{content}}.
\]

The next section details the input representations $z_0$ used by each
network.



\section{Entity Relatedness Ranking}

\subsection{Content-based representation learning}

To obtain content representations, we leverage both the textual document
(\textit{word}-based) and the link profile (\textit{entity}-based) of
each entity, as introduced in Section~\ref{sec:prel}.  
Given the large word and entity vocabularies, we employ the
\textit{word hashing} technique of \cite{huang2013learning}, which maps
each token to a set of character trigrams.  
Let $\mathcal{V}$ denote the trigram vocabulary and
$\mathsf{E}:\mathcal{V}\!\rightarrow\!\mathbb{R}^m$ the embedding
function.  
We also learn a global importance weight
$\mathsf{w}:\mathcal{V}\!\rightarrow\!\mathbb{R}_{\ge 0}$.

For an entity $e$, let $e_w = (v_1,\ldots,v_{n_e})$ denote the sequence
of hashed trigrams extracted from its document.  
We represent $e$ by a weighted compositional embedding:
\[
\phi_{\mathrm{word}}(e)
=
\sum_{j=1}^{n_e}
\mathsf{w}(v_j)\,\mathsf{E}(v_j)
\;\in\mathbb{R}^m.
\]
This formulation corresponds to a linear bag-of-subword encoder and is
equivalent to the expected embedding of a token drawn proportionally to
its trigram weights.

For the entity-based representation, we analogously define
$e_{\mathrm{ent}} = (u_1,\ldots,u_{k_e})$ as the hashed tokens obtained
from the surface forms of entities linked from $e$, and compute:
\[
\phi_{\mathrm{ent}}(e)
=
\sum_{j=1}^{k_e}
\mathsf{w}(u_j)\,\mathsf{E}(u_j).
\]
The final content representation is their concatenation:
\[
\phi_{\mathrm{content}}(e)
=
[\phi_{\mathrm{word}}(e);
 \phi_{\mathrm{ent}}(e)]
\in\mathbb{R}^{2m}.
\]
For a triplet $(e_s,e_{(+)},e_{(-)})$, the concatenated content
representations serve as the input to the content sub-network.

\subsection{Graph-based representation}

For graph-based representations, we follow the DeepWalk framework
\cite{perozzi2014deepwalk}.  
Let $G=(V,E)$ be the entity graph, where $V$ is the set of entities and
edges correspond to hyperlinks.  
DeepWalk generates random walks
$\mathbb{S}_{e}=(v_1,\ldots,v_L)$ from each entity $e$ and optimizes a
Skip-gram objective:
\[
\max_{\mathsf{G}}
\sum_{e\in V}
\sum_{v_j\in \mathbb{S}_e}
\sum_{v_k\in \mathcal{N}(v_j)}
\log 
\mathbb{P}\big(
v_k \mid \mathsf{G}(v_j)
\big),
\]
where $\mathsf{G}:V\!\rightarrow\!\mathbb{R}^d$ is the learned graph
embedding and $\mathcal{N}(v_j)$ is the context window.  
This yields graph-aware embeddings capturing co-occurrence structure
under random walks.

Given two entities $e_s$ and $e_t$, let
\[
\mathbb{C}_{e_s} = \{\mathsf{G}(v): v\in\mathbb{S}_{e_s}\},\qquad
\mathbb{C}_{e_t} = \{\mathsf{G}(u): u\in\mathbb{S}_{e_t}\}
\]
denote their graph-context embeddings.  
For every pair $(x,y)\in(\mathbb{C}_{e_s},\mathbb{C}_{e_t})$, we compute
the cosine similarity:
\[
\mathrm{sim}(x,y)
=
\frac{\langle x,y\rangle}{
\lVert x\rVert\,\lVert y\rVert}.
\]

Following \cite{guo2016deep}, we discretize these similarities into
$B$ fixed histogram bins.  
Let $H_b$ be the count of similarities falling into bin $b$.  
The interaction vector is:
\[
\phi_{\mathrm{graph}}(e_s,e_t)
=
\big[
\log(1+H_1),\;
\ldots,\;
\log(1+H_B)
\big]
\in\mathbb{R}^B.
\]
This histogram encodes soft matching of graph neighborhoods, analogous
to classical link-based relatedness measures~\cite{witten2008effective},
but operating at the embedding level.

\subsection{Temporal Representation via Time-Weighted Convolution}

To learn representations from temporal signals, we treat each entity’s
multivariate time series as a sequence of $T$ aligned observations.
Our goal is to embed such sequences into a Euclidean space in which
temporally similar entities lie close together.  
We employ a 1-D convolutional architecture that extracts local temporal
patterns and then reweights these patterns according to their temporal
proximity to the prediction time.  
The architecture consists of: (i) a temporal convolution layer,
(ii) batch normalization, (iii) a time-decay weighting mechanism,
and (iv) a fully connected projection layer.

\paragraph{Temporal convolution.}
Let $X = (x_1, \ldots, x_T)$ denote an entity’s time series, where
$x_t \in \mathbb{R}^{D}$ is a $D$-dimensional feature vector at time $t$
(e.g., a scalar popularity signal or a small feature bundle).
A 1-D convolution with kernel width $w$ applies a filter
$\mathbf{W} \in \mathbb{R}^{w \times D}$ and bias $b \in \mathbb{R}$ to
each contiguous window $(x_{t-w+1}, \ldots, x_t)$:
\[
q_t = \langle \mathbf{W}, X_{t-w+1:t} \rangle + b,
\qquad t = w,\ldots,T .
\]
Batch normalization and a ReLU activation yield
\[
h_t = \mathrm{ReLU}\!\left(\mathrm{BN}(q_t)\right),
\qquad t = w,\ldots,T .
\]
The resulting sequence
$\mathbf{h} = (h_w, h_{w+1}, \ldots, h_{T})$
captures local temporal patterns (bursts, surges, and short-term trends).

\paragraph{Time-decay weighting.}
Not all temporal patterns should contribute equally:
patterns closer to the prediction time are intuitively more relevant.
We therefore introduce a deterministic, strictly positive decay function
$A_t$ that assigns higher weight to features extracted from recent time
steps.

For the convolution output index $t$, denote its temporal distance from
the prediction point by $\Delta_t = T - t$.
We define:
\[
A_t = \frac{1}{(1 + \Delta_t)^{\alpha}},
\qquad \alpha > 0,
\quad t = w,\ldots,T .
\]
The weighted convolutional representation is then:
\[
\tilde{h}_t = A_t \cdot h_t,
\qquad t = w,\ldots,T .
\]
This mechanism is a deterministic ``soft focusing'' operation:  
unlike learned attention, the weights are fixed by temporal proximity,
providing an interpretable, monotonic recency bias.

\paragraph{Projection.}
Finally, the weighted sequence is flattened and passed through a
nonlinear fully connected layer:
\[
z = \sigma\!\left(\mathbf{W}_{\mathrm{fc}}
\,[\tilde{h}_w;\tilde{h}_{w+1};\ldots;\tilde{h}_T] + \mathbf{b}_{\mathrm{fc}}\right),
\]
which yields the final temporal embedding $z \in \mathbb{R}^{d}$.
Only the last convolutional layer is time-weighted, though deeper stacks
of convolutional layers can be used.

This design cleanly separates (i) local temporal pattern extraction via
convolution, and (ii) global temporal relevance modulation via a
principled, monotonic decay, yielding an interpretable and robust
temporal representation.

\begin{figure}[t]
\centering
\includegraphics[angle=270,origin=c,width=0.8\columnwidth]{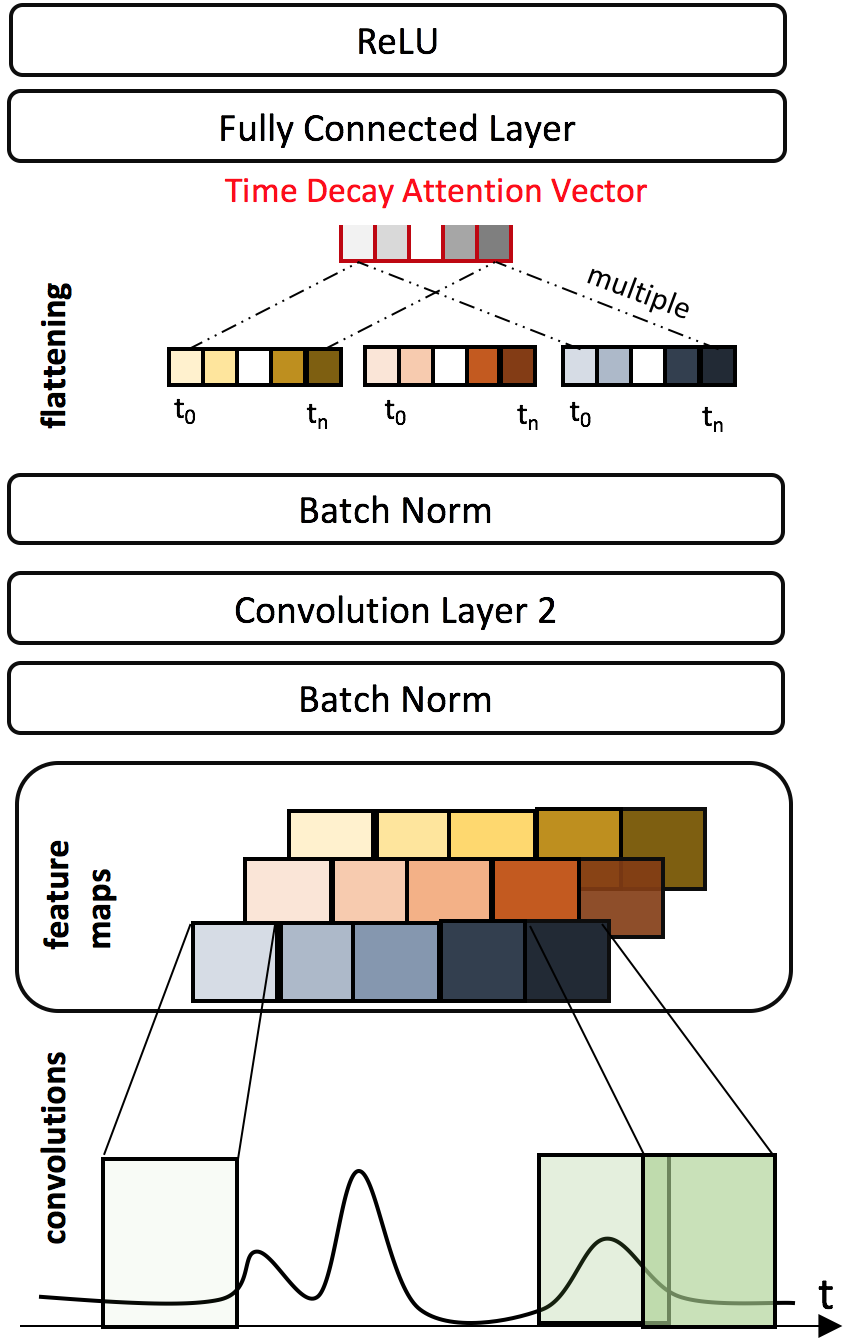}
\caption{Architecture of the temporal convolution module.
Local temporal patterns are extracted by a 1-D convolution and then
modulated by a monotonic time-decay weighting before projection into the
final temporal embedding space.}
\label{fig:acnn}
\end{figure}


\subsection{Learning and Optimization}

Our model learns a pairwise preference function over entity candidates.
For each training instance $i$ consisting of a source entity $e_s$ and a
positive--negative pair $(e_{(+)}, e_{(-)})$, the model produces a score
$f_{t(i)}(e_s, e)$ for each candidate.  
The probability that $e_{(+)}$ should be ranked above $e_{(-)}$ is
defined through a logistic link applied to the score difference:
\[
\bar{y}_i
=
\sigma\!\big(
f_{t(i)}(e_s, e_{(+)})
-
f_{t(i)}(e_s, e_{(-)})
\big).
\]

To obtain supervision, we construct soft preference targets from
navigation statistics:
\[
\tilde{P}_i
=
\frac{
y^{t(i)}_{\{e_s, e_{(+)}\}}
}{
y^{t(i)}_{\{e_s, e_{(+)}\}}
+
y^{t(i)}_{\{e_s, e_{(-)}\}}
},
\]
which provide a smoothed estimate of how often users prefer
$e_{(+)}$ over $e_{(-)}$ when navigating from $e_s$ at time $t(i)$.

We train the network using the regularized cross-entropy objective:
\[
L
=
-\frac{1}{N}
\sum_{i=1}^{N}
\Big(
\tilde{P}_i \log \bar{y}_i
+
(1 - \tilde{P}_i)\log(1 - \bar{y}_i)
\Big)
+
\lambda \lVert \theta \rVert_2^2,
\]
where $\theta$ denotes all model parameters.
This loss is a smooth pairwise surrogate for ranking and provides
calibrated probability estimates, enabling the model to distinguish fine
differences in temporal relatedness. Parameters are optimized using Adam \cite{kingma2014adam}.


%% file: clickstream.tex
\textbf{Clickstream.}
\label{clickstreamsec}
For entity navigation, we use the clickstream dataset
generated from the Wikipedia webserver logs from February until September, 2016. These datasets contain
an accumulation of transitions between two Wikipedia articles with their respective counts on a monthly basis. We study only actual pages (e.g. excluding disambiguation or redirects). In the following, we provide the first analysis of the clickstream data to gain insights into the temporal dynamics of the entity collective attention in Wikipedia.
\begin{figure*}[ht!]
\centering
\begin{subfigure}[t]{0.3\textwidth}
   \includegraphics[width=1\linewidth]{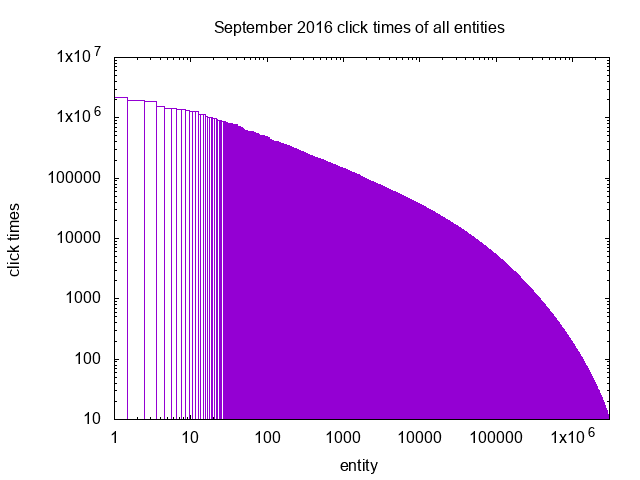}
   \caption{Click times distribution}
   \label{fig:ctdist}
\end{subfigure}
\centering
   \begin{subfigure}[t]{0.3\textwidth}
   \includegraphics[width=1\linewidth]{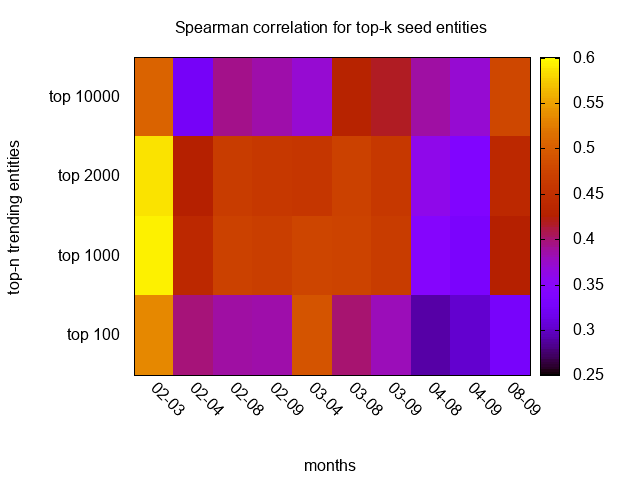}
   \caption{Correlation of top-k entities}
   \label{fig:topkheat} 
\end{subfigure}
\begin{subfigure}[t]{0.3\textwidth}
   \includegraphics[width=1\linewidth]{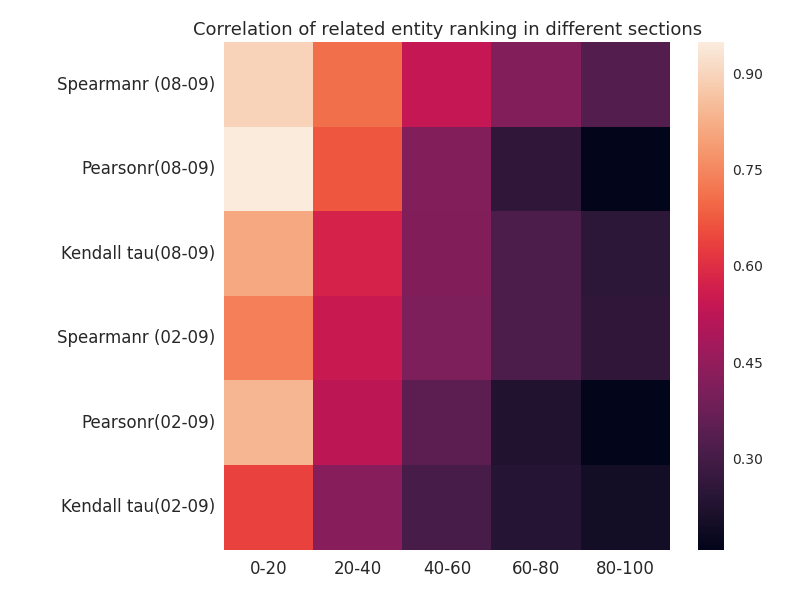}
   \caption{Correlation by \# of navigations}
   \label{fig:corrheat}
\end{subfigure}
\caption{Click (navigation) times distribution and ranking correlation of entities in September 2016.}
\label{fig:singles}
\end{figure*}

\begin{table}[]
\centering
\caption{Statistics on the dynamic of clickstream, $e_{s}$ denote source entities, $e_{t}$ related entities.}
\label{tab:clickstat}
\vspace{-0.2cm}
\def\arraystretch{1.0} \small
\scriptsize{\begin{tabular}{p{1cm}|p{0.8cm}p{1.4 cm}p{1.2cm}p{1cm}}
\toprule
                 & \textbf{\% new $e_{s}$} & \textbf{\% with new $e_{t}$} & \textbf{\% w. new $e_{t}$ in top-30} & \textbf{\# new $e_{t}$ (avg.)} \\ \hline
\textbf{08-2016} & 24.31                                                 & 71.18                                                        & 15.54                                             & 18.25                         \\ \hline
\textbf{04-2016} & 30.61                                                 & 66.72                                                        & 53.44                                             &   42.20                       \\ \bottomrule
\end{tabular}}
\vspace{-0.2cm}
\end{table}

Figure~\ref{fig:ctdist} illustrates the distribution of entities by click frequencies, and the correlation of top popular entities (measured by total navigations) across different months is shown in Figure~\ref{fig:topkheat}. In general, we observe that the user navigation activities in the top popular entities are very dynamic, and changes substantially with regard to time. Figure~\ref{fig:corrheat} visualizes the dynamics of related entities toward different ranking sections  (e.g., from rank 0 to rank 20) of different months, in terms of their correlation scores. It can be interpreted that the entities that stay in top-20 most related ones tend to be more correlated than entities in bottom-20 when considering top-100 related entities. 

As we show in Table~\ref{tab:clickstat}, there are 24.31\% of entities in top-10,000 most active entities of September 2006 do not appear in the same list the previous month. And 30.61\% are new compared with 5 months before. In addition, there are 71\% of entities in top-10,000 having navigations to new entities compared to the previous month, with approx. 18 new entities are navigated to, on average. Thus, the datasets are naturally very dynamic and sensitive to change. The substantial amount of missing \textit{past} click logs on the \textbf{newly-formed relationships} also raises the necessity of an dynamic measuring approach. 

%% file: experiment.tex
\section{Experiments}
\subsection{Dataset}
\label{ref:exp:dataset}
To recap from Section~\ref{clickstreamsec}, we use the click stream datasets in 2016. We also use the corresponding Wikipedia article dumps, with over 4 million entities represented by actual pages. Since the length of the content of an Wikipedia article is often long, in this work, we make use of only its \textbf{abstract} section. To obtain temporal signals of the entity, we use page view statistics
of Wikipedia articles and aggregate the counts by month. We fetch the data from June, 2014 up until the studied time, which results in the length of 27 months. 

\textbf{Seed entities and related candidates.} To extract popular and trending entities, we extract from the clickstream data the top 10,000 entities based on the number of navigations from major search engines (\textit{Google} and \textit{Bing)}, at the studied time.  Getting the subset of related entity candidates --for efficiency purposes-- has been well-addressed in related work~\cite{guo2014robust,Ponza:2017:TFC:3132847.3132890}. In this work, we do not leverage a method and just assume the use of an appropriate one. In the experiment, we resort to choose only candidates which are visited from the seed entities at studied time. We filtered out entity-candidate pairs with too few navigations (less than 10) and considered the top-100 candidates.

\begin{table}[t]
    \centering
    \tabcolsep=0.08cm
    \caption{Statistics of the dataset.}
     \scriptsize{
    \begin{tabular}{l|l}
    \tabcolsep=0.08cm
   
      & Counts \\
    \hline
    Total seed entities & $10,000$  \\ 
    Total entities & $1,420,819$ \\
    Candidate per entities (avg.) & $142$ \\
    \midrule
    Training seed entities & $8,000$\\
    Dev. seed entities & $1,000$\\
    Test seed entities & $1,000$\\
    \midrule
    Training pairs & $100,650K$\\
    Dev. pairs & $12,420K$\\
    Test pairs & $12,590K$\\
    \bottomrule
    \end{tabular}}
    \vspace{-0.1cm}
    \label{tab:dataset}
\end{table}

%

%

\subsection{Models for Comparison}
In this paper, we compare our models against the following
baselines.

\textbf{ Wikipedia Link-based} (WLM):~\newcite{witten2008effective} proposed a low-cost measure of semantic
relatedness based on Wikipedia entity graph, inspired
by Normalized Google Distance. 

\textbf{DeepWalk} (DW): DeepWalk~\cite{perozzi2014deepwalk}
learned representations of vertices in a graph with a random
walk generator and language modeling. We chose not to compare with the matrix factorization approach in~\cite{zhao2015representation}, as even though it allows the incorporation of different relation types (i.e., among entity, category and word), the iterative computation cost over large graphs is very expensive. When consider only entity-entity relation, the performance is reported rather similar to DW. 

\textbf{Entity2Vec Model} (E2V): or entity embedding learning using Skip-Gram~\cite{mikolov2013distributed} model. E2V utilizes textual information to capture latent
word relationships. Similar to~\newcite{zhao2015representation,Ni:2016:SDR:2835776.2835801}, we use Wikipedia articles as training corpus to learn word vectors and reserved hyperlinks between
entities.

\textbf{ParaVecs} (PV): ~\newcite{le2014distributed,dai2015document} learned document/entity vectors via the distributed memory (\textbf{ParaVecs-DM}) and distributed bag of words (\textbf{ParaVecs-DBOW}) models, using hierarchical softmax. We use Wikipedia articles as training corpus to learn entity vectors. 

\textbf{RankSVM}: ~\newcite{ceccarelli2013learning} learned entity relatedness from a set of 28 \textbf{handcrafted features}, using the traditional learning-to-rank method, RankSVM. We put together additional well-known temporal features~\cite{kanhabua2014triggers,zhang2016probabilistic} (i.e., time series cross correlation, trending level and predicted popularity based on \textit{page views}) and report the results of the extended feature set.

For our approach, we tested different combinations of \textit{content} (denoted as $\mathbf{Content_{Emb}}$), \textit{graph}, ($\mathbf{Graph_{Emb}}$) and \textit{time} (\textbf{TS-CNN-Att}) networks. We also test the \textit{content} and \textit{graph} networks with \textbf{pretrained} entity representations (i.e., ParaVecs-DM and DeepWalk).



\subsection{Experimental Setup}
\vspace*{-0.1cm}
\textbf{Evaluation procedures.} The time granularity is set to months. The studied time $t_{n}$ of our experiments is September 2016. From the seed queries, we use 80\% for training, 10\% for development and 10\% for testing, as shown in Table~\ref{tab:dataset}. Note that, for the \textbf{time-aware} setting and to avoid leakage and bias as much as possible, the data for training and development (including supervision) are up until time $t_{n}-1$. In specific, for content and graph data, only $t_{n}-1$ is used.

\textbf{Metrics.} We use 2 correlation coefficient methods, Pearson and Spearman, which have been used often throughout literature, cf.~\cite{dallmann2016extracting,Ponza:2017:TFC:3132847.3132890}. The Pearson index focuses on the difference between predicted-vs-correct relatedness scores, while Spearman focuses on the ranking order among entity pairs. Our work studies on the strength of the dynamic relatedness between entities, hence we focus more on Pearson index. However, traditional correlation metrics do not consider the positions in the ranked list (correlations at the top or bottom are treated equally). For this reason, we adjust the metric to consider the rankings at specific top-k positions, which consequently can be used to measure the correlation for only top items in the ranking (based to the ground truth). 
In addition, 
we use Normalized Discounted Cumulative Gain (NDCG) measure to evaluate the recommendation tasks.

\textbf{Implementation details.} All neural models are implemented in TensorFlow. Initial learning
rate is tuned amongst \{1.e-2, 1.e-3, 1.e-4, 1.e-5\}. The batch size is tuned
amongst \{50, 100, 200\}. The weight
matrices are initialized with samples from the uniform
distribution~\cite{glorot2010understanding}. Models are trained for maximum 25 epochs. The hidden layers for each network are among \{2, 3, 4\}, while for hidden nodes are \{128, 256, 512\}. Dropout rate is set from \{0.2, 0.3, 0.5\}. The pretrained DW is empirically set to 128 dimensions,  and 200 for PV. For CNN, the filter number are in \{10, 20, 30\}, window size in ~\{4, 5, 6\}, convolutional layers in~\{1, 2, 3\} and decay rate $\alpha$ in \{1.0, 1.5,$\cdots$,7.5\}. 2 conv- layers with window size 5 and 4, number of filters of 20 and 25 respectively are used for decay hyperparameter analysis. 


\begin{table*}[ht!]
\centering
\caption{Performance of different models on task (1) Pearson, Spearman's $\rho$ ranking correlation, and task (2) recommendation (measured by nDCG). Bold and underlined numbers indicate best and second-to-best results. $\mp$ shows statistical significant over WLM ($p<0.05$).}
\label{tab:exp}
\small{
\begin{tabular}{lllllllllllll}
\toprule
\multirow{2}{*}{}                                         & \multicolumn{1}{l|}{\multirow{2}{*}{\textbf{Model}}} & \multicolumn{4}{c|}{\textbf{Pearson $\times 100$ }}                                                                     & \multicolumn{1}{l|}{\textbf{$\rho \times 100$}} & \multicolumn{3}{l|}{\textbf{nDCG} (proxy)}                                            & \multicolumn{3}{c}{\textbf{nDCG} (human)}                                             \\ \cline{3-13} 
                                                          & \multicolumn{1}{l|}{}                                & \multicolumn{1}{l|}{@10} & \multicolumn{1}{l|}{@30} & \multicolumn{1}{l|}{@50} & \multicolumn{1}{l|}{all} & \multicolumn{1}{l|}{all}               & \multicolumn{1}{l|}{@3} & \multicolumn{1}{l|}{@10} & \multicolumn{1}{l|}{@20} & \multicolumn{1}{l|}{@3} & \multicolumn{1}{l|}{@10} & \multicolumn{1}{l}{@20} \\ \hline
{\multirow{6}{*}{\rotatebox[origin=c]{90}{\textbf{Baselines}}}} &  WLM         & 27.6                    & 28.3                   & 24.0                    & 19.4                    &    12.1                                & 0.63                   & 0.59                    & 0.62                    &          0.50          & 0.46                   &                    0.52 \\
                                  & RankSVM                                              & 28.5                & 34.7                & 31.4               & 20.7               & 27.5                              & 0.65 & 0.61 & 0.64                    &                        0.52 &             0.61             &    0.65                      \\
                                   & Entity2Vec                                           & 18.6   &   22.0             & 21.8                & 20.5                & 18.7                             & 0.62 & 0.60                & 0.61                &    0.54                     &                         0.53 &         0.54                 \\
                                 & DeepWalk                                             & 31.3                    & 30.9                    & 21.4                    & 17.6                    & 10.1                                  &    0.41                & 0.43                    & 0.47                    &  0.34&0.38                     &0.45                     \\
                                   & ParaVecs-DBOW                                        & 18.6                    & 22.0                    & 21.8                    & 20.5 & 16.0                                  & 0.62 & 0.60                    & 0.61                    &     0.50               &                    0.50 &  0.55                   \\
                                    & ParaVecs-DM                                          & 19.0 & 23.0                    & 23.2                    & 22.3 & 18.3                                  & 0.66                   & 0.63                    & 0.63                   &                   0.49 &          0.52           &     0.58                \\ 
\midrule
\multirow{8}{*}{\rotatebox[origin=c]{90}{\textbf{Model Ablation}}}                                         & TS-CNN                                          & 51.9 & 51.0                    & 43.0                    & 35.8                    & 26.5                                  & 0.41                   & 0.43                    & 0.47 &0.40                    &  0.43                   & 0.48 \\
                                                          & TS-CNN-Att (\textbf{Base}) & 57.9                    & 49.7                    & 44.7                   & 37.1                 & 24.9                                  &0.43 & 0.44 & 0.49                    &  0.38                 & 0.45 & 0.50  \\
                                                          & Base+PV & \underline{60.6}                    & 44.2                    & 41.4                    & 36.4                    & 11.2 & 0.41                   & 0.43                    & 0.47                    &0.49  &0.51                     &    0.55                \\
                                                          & Base+DW                                         & 43.5                    & 36.5                    & 35.7                    & 32.7                    & 31.0                                  & 0.44                   & 0.48                    & 0.53                    &     0.47               &     0.51                & 0.52  \\
                                                          & Base+PV+DW                                      &56.9 & 46.1                    & 43.4                    & 32.9                   & 28,4                                  & 0.41                   & 0.44                 & 0.48                    &  0.49                  &       0.54              &0.57  \\
\cmidrule{2-13}
& $Content_{Emb}$+$Graph_{Emb}$ & 48.9 &                      40.1&             49.9      &           37.5         &         27.9                        &        0.67            &           0.62          &   0.70                  &                   0.61 &          0.69           & 0.65 \\                                                         
                                                               & Base+$Content_{Emb}$                                                                &\textbf{67.1} &                      \underline{54.2}&             \textbf{53.4}       &           \underline{43.7}          &         26.5                         &        0.67            &           0.69          &   0.71                  &                   0.61 &          0.72           & 0.74 \\
& Base+$Graph_{Emb}$ & 55.2&    50.2                 &  41.3&                    31.5&                \underline{35.5}                   & \underline{0.71} &\underline{0.75}   & \underline{0.78} &                   \underline{0.65}$^{\mp}$   &        \underline{0.78}$^{\mp}$ & \underline{0.81}$^{\mp}$   \\
\midrule  & \textbf{Trio}                                                 &     58.6                     &        \textbf{54.3}                  &     \underline{50.2}                     &       \textbf{45.4}                   &            \textbf{43.5}                            &      \textbf{0.75}                   &                        \textbf{ 0.78} &           \textbf{0.83}               &    \textbf{0.74}$^{\mp}$                   &                         \textbf{0.82}$^{\mp}$ &         \textbf{0.85}$^{\mp}$               \\ 
\bottomrule
\end{tabular}
}

\end{table*}
\vspace*{-0.1cm}
\subsection{Experimental Tasks}
We evaluate our proposed method in two different scenarios: (1) Relatedness ranking and (2) Entity recommendation. The first task evaluates how well we can \textbf{mimic} the ranking via the entity navigation. Here we use the raw number of navigations in Wikipedia clickstream. 
The second task is formulated as: \textit{given an entity, suggest the top-k most related entities to it right now}. Since there is no standard ground-truth for this temporal task, we constructed two relevance ground-truths. The \textbf{first} one is the \textit{proxy} ground-truth, with relevance grade is \textit{automatically} assigned from the (top-100) most navigated target entities. The graded relevance score is then given as the \textit{reversed} rank order. For this, all entities in the test set are used. The \textbf{second} one is based on the human judgments with 5-level graded relevance scale, i.e., from 4 - highly relevant to 0 - not (temporally) relevant. Two human experts evaluate on the subset of 20 entities (randomly sampled from the test set), with 600 entity pairs (approx. 30 per seed, using \textit{pooling} method). The ground-truth size is comparable the widely used ground-truth for \textit{static} relatedness assessment, KORE~\cite{hoffart2012kore}.  The Cohen's Kappa agreement is 0.72. Performance of the best-performed models on this dataset is then tested with paired \textit{t}-test against the WLM baseline.
\subsection{Results on Relatedness Ranking}
We report the performance of the relatedness ranking on the left side of Table~\ref{tab:exp}, with the Pearson and Spearman metrics. Among existing baselines, we observe that link-based approaches i.e., WLM and DeepWalk perform better than others for top-k correlation. Whereas, temporal models yield substantial improvement overall. Specifically, the TS-CNN-Att performs better than the no-attention model in most cases, improves 11\% for Pearson@10, and 3\% when considering the total rank. Our trio model performs well overall, gives best results for total rank. The duo models (combine base with either pretrained DW or PV) also deliver improvements over the sole temporal ones. We also observer additional gains while combining of temporal base with pretrained DW and PV altogether. 

\begin{figure*}[t!]
\centering
%
\begin{subfigure}[t]{0.3\textwidth}
   \includegraphics[width=1\linewidth]{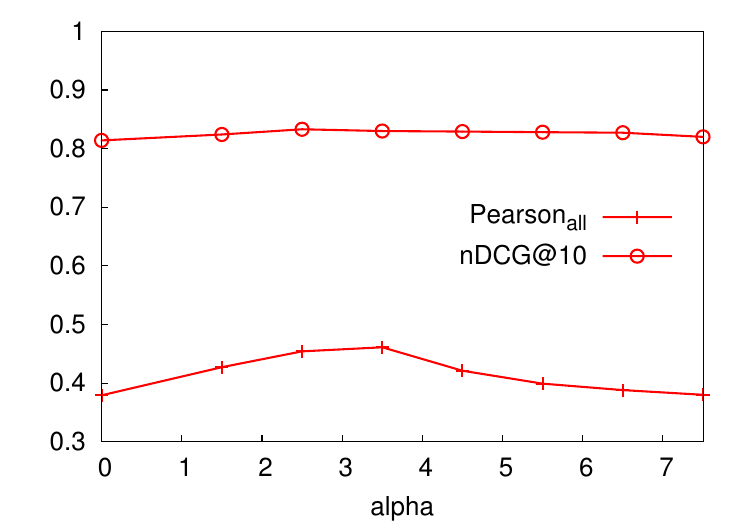}
   \caption{Decay parameter for time-series embedding.}
   \label{fig:decp}
\end{subfigure}
\centering
   \begin{subfigure}[t]{0.3\textwidth}
   \includegraphics[width=1\linewidth]{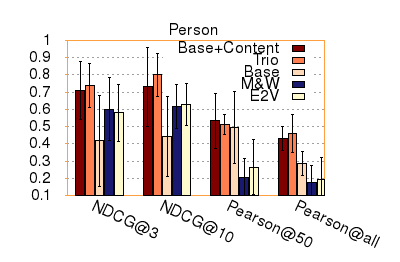}
   \caption{Model performances for \textit{person}-type entities.}
   \label{fig:person} 
\end{subfigure}
\begin{subfigure}[t]{0.3\textwidth}
   \includegraphics[width=1\linewidth]{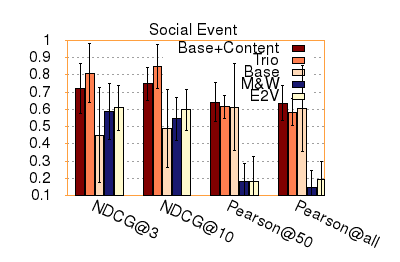}
   \caption{Model performances for \textit{social event}-type entities.}
   \label{fig:event}
\end{subfigure}
\caption{Performance results for variation of decay parameter and different entity types.}
\label{fig:resultmixed}
\end{figure*}

\begin{figure}[h]
\centering
		\includegraphics[width=0.7\columnwidth]{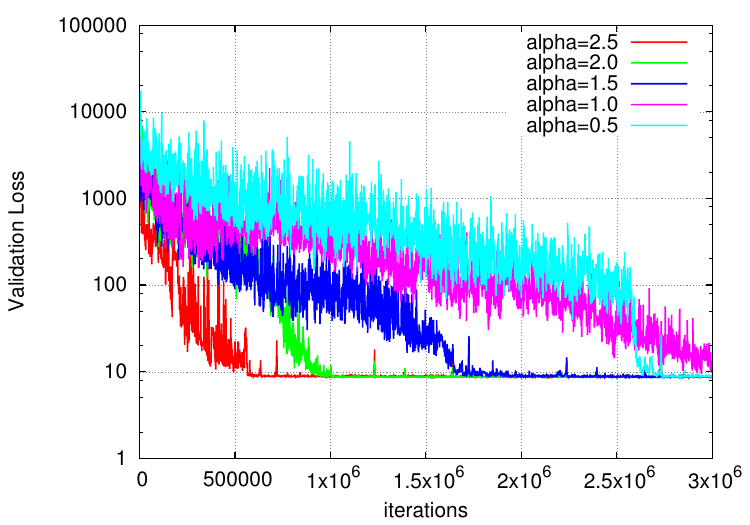}\vspace{-0.1cm}
		\caption{Convergence of decay parameters.}		
		\label{fig:convergence}
\end{figure}
\subsection{Results on Entity Recommendation}
Here we report the results on the nDCG metrics. Table~\ref{tab:exp} (right-side) demonstrates the results for two ground-truth settings (proxy and human). We can observe the good performance of the baselines for this task over conventional temporal models, significantly for \textit{proxy} setting. It can be explained that, `static' entity relations are ranked high in the non time-aware baselines, hence are still rewarded when considering a fine-grained grading scale (100 level). The margin becomes smaller when comparing in \textit{human} setting, with the standard 5-level scale. All the models with pretrained representations perform poorly. 
It shows that for this task, early interaction-based approach is more suitable than purely based on representation.

\begin{table}[]
\centering
\scriptsize{
\caption{Different top-k rankings for entity \textit{Kingsman: The Golden Circle}. Italic means irrelevance.} 
\label{fig:anecdotic}
\begin{tabular}{cccccc}

\toprule
\multicolumn{4}{c}{\textbf{Models}}                                                                \\
\textbf{PV-DM}                     & \textbf{TS-CNN-Att}           & \textbf{Temp+PV}    & \textbf{Trio} \\
\midrule
Secret Service    & Halle Berry                & \textit{Elton John} & Mark Strong   \\
\textit{Spider-Man} & \textit{X-Men} & Taron Egerton       & Jeff Bridges  \\
Taron Egerton                   & Jeff Bridges               & Edward Holcroft     & Julianne More \\
\bottomrule
\end{tabular}}
\end{table}
\subsection{Additional Analysis}
We present an anecdotic example of top-selected entities for \textsf{Kingsman: The Golden Circle} in Table~\ref{fig:anecdotic}. While the content-based model favors old relations like the preceding movies, TS-CNN puts popular actress \textsf{Halle Berry} or the recent released \textsf{X-men: Apocalypse} on top. The latter is not ideal as there is not a solid relationship between the two movies. One implication is that the two entities are ranked high is more because of the popularity of themself than the strength of the relationship toward the source entity. The Trio model addresses the issue by taking other perspectives into account, and also balances out the recency and long-term factors, gives the best ranking performance.

\textbf{Analysis on decay hyper-parameter.}
We give a study on the effect of decay parameter on performance. Figure~\ref{fig:decp} illustrates the results on $Pearson_{all}$ and nDCG@10 for the \textit{trio} model. It can be seen that while nDCG slightly increases, Pearson score peaks while $\alpha$ in the range $[1.5,3.5]$. Additionally, we show the convergence analysis on $\alpha$ for TS-CNN-Att in Figure~\ref{fig:convergence}. Bigger $\alpha$ tends to converge faster, but to a significant higher loss when $\alpha$ is over 5.5 (omitted from the Figure). 


\textbf{Performances on different entity types.}
We demonstrate in Figures~\ref{fig:person} and ~\ref{fig:event} the model performances on the \textit{person} and \textit{event} types. WLM performs poorer for the latter, that can be interpreted as link-based  methods tend to slowly adapt for recent trending entities. The temporal models seem to capture these entites better.

%% file: conclusion.tex
\vspace{-0.25cm}
\section{Conclusion}
\vspace{-0.25cm}
In this work, we presented a trio neural model to solve the dynamic entity relatedness ranking problem. The model jointly learns rich representations of entities from textual content, graph and temporal signals. We also propose an effective CNN-based attentional mechanism for learning the temporal representation of an entity. Experiments on ranking correlations and top-$k$ recommendation tasks demonstrate the effectiveness of our approach over existing baselines. For future work, we aim to incorporate more temporal signals, and investigate on different `trainable' attention mechanisms to go beyond the time-based decay, for instance by incorporating latent topics. 

%% file: acknowledgements.tex
\vspace{0.2cm}

\noindent\textbf{Acknowledgments.}
This work is funded by the ERC Advanced Grant
ALEXANDRIA (grant no. 339233). We thank the reviewers for the
suggestions on the content and structure of the paper.